\documentclass[aps, prb, superscriptaddress, twocolumn, letterpaper, 10pt, amsfonts, amsmath, amssymb]{revtex4-2}

\usepackage{graphicx}
\usepackage{hyperref} % For hyperlinks in the PDF
\usepackage{xcolor}
\usepackage{soul}
\hypersetup{
    colorlinks,
    linkcolor={blue!80!black},
    citecolor={blue!80!black},
    urlcolor={blue!80!black}
}
\usepackage{physics}
\usepackage{placeins}

\usepackage{siunitx}
\usepackage{chemformula}

\makeatletter
\def\maketitle{
\@author@finish
\title@column\titleblock@produce
\suppressfloats[t]}
\makeatother

\begin{document}

\title{Interference and parity blockade in transport through a Majorana box}
\author{Maximilian Nitsch}
\affiliation{Division of Solid State Physics and NanoLund, Lund University, S-22100 Lund, Sweden}
\author{Rubén Seoane Souto}
\affiliation{Division of Solid State Physics and NanoLund, Lund University, S-22100 Lund, Sweden}
\affiliation{Center for Quantum Devices, Niels Bohr Institute, University of Copenhagen, DK-2100 Copenhagen, Denmark}
\author{Martin Leijnse}
\affiliation{Division of Solid State Physics and NanoLund, Lund University, S-22100 Lund, Sweden}
\affiliation{Center for Quantum Devices, Niels Bohr Institute, University of Copenhagen, DK-2100 Copenhagen, Denmark}

\date{\today}

\begin{abstract}
A Majorana box – two topological superconducting nanowires coupled via a trivial superconductor – is a building block in devices aiming to demonstrate nonabelian physics, as well as for topological quantum computer architectures. We theoretically investigate charge transport through a Majorana box and show that current can be blocked when two Majoranas couple to the same lead, fixing their parity. In direct analogy to Pauli spin blockade in spin qubits, this parity blockade can be used for fast and high-fidelity qubit initialization and readout, as well as for current-based measurements of decoherence times. Furthermore, we demonstrate that transport can distinguish between a clean Majorana box and a disordered box with additional unwanted Majorana or Andreev bound states.
%We show that the connection to the leads induces a Lamb-shift in the many-body eigenenergies which makes it possible to experimentally distinguish between a clean Majorana box qubit and a disordered box with additional unwanted Majorana or Andreev bound states.
%The parity blockade in a standard Majorana box relies on fine tuning of tunnel couplings and we therefore propose a single-sided box with a more robust blockade.
\end{abstract}

\maketitle

\emph{Introduction.}
Topological $p$-wave superconductors host Majorana bound states (MBSs)~\cite{Kitaev_2001,NayakReview,Alicea_RPP2012,LeijnseReview,AguadoReview,BeenakkerReview_20} at edges and defects, which have nonlocal and nonabelian properties. Semiconductor nanowires are one of the most promising systems for creating and detecting MBSs, where a combination of spin-orbit coupling, proximity-induced superconductivity, and external magnetic field can lead to $p$-wave superconductivity~\cite{Oreg_PRL2010,Lutchyn_PRL2010}. By now, many experiments have observed zero-bias conductance peaks, consistent with MBSs at the nanowire ends (see Refs.~\cite{Mourik_science2012,deng2012anomalous,finck2013anomalous,deng2016majorana,Nichele_PRL2017,lutchyn2018majorana} for a few examples, similar results have been obtained also in other MBS platforms). However, nontopological states provide an alternative explanation for most of the experimental observations~\cite{Prada_PRB2012,Kells_PRB12,Moore_PRB18,Awoga_PRL2019,Vuik_SciPost19,Pan_PRR20,hess2021local}.

A measurement of the nonabelian properties of MBSs is still missing, but would provide definite evidence of a topological superconducting phase, constituting at the same time a first step towards topological quantum computing. One promising path towards a demonstration of nonabelian physics uses repeated measurements of MBS pairs to perform topologically protected qubit operations~\cite{Bonderson_PRL2008}, with a possibility to move towards a scalable quantum computer platform~\cite{Vijay_PRX2015,Plugge_PRB2016,Karzig_PRB2017}. A simple building block for these technologies is the Majorana box qubit~\cite{Plugge_NJP2017}, where a qubit is encoded in four MBSs with overall parity fixed by a large charging energy. Qubit readout can be done by charge sensing of a quantum dot coupled to two MBSs~\cite{munk2020parity,steiner2020readout,Smith_PRXQ2020,Schulenborg_PRB2021}, or by measuring the interference of cotunneling currents when the box is connected to external leads~\cite{fu2010electron}. Furthermore, coupling the Majorana box to four leads enables measurements of the topological Kondo effect~\cite{Beri_PRL12,Galpin_PRB2014,Buccheri_book20},
and networks of coupled Majorana boxes exhibit additional interesting transport physics~\cite{Beri_PRL13,Altland_PRL2013,Herviou_PRB2016,Michaeli_PRB2017,Vayrynen_PRR2020}. 
\begin{figure}[h]
	\begin{center}
		\includegraphics[angle=0,width=0.45\textwidth]{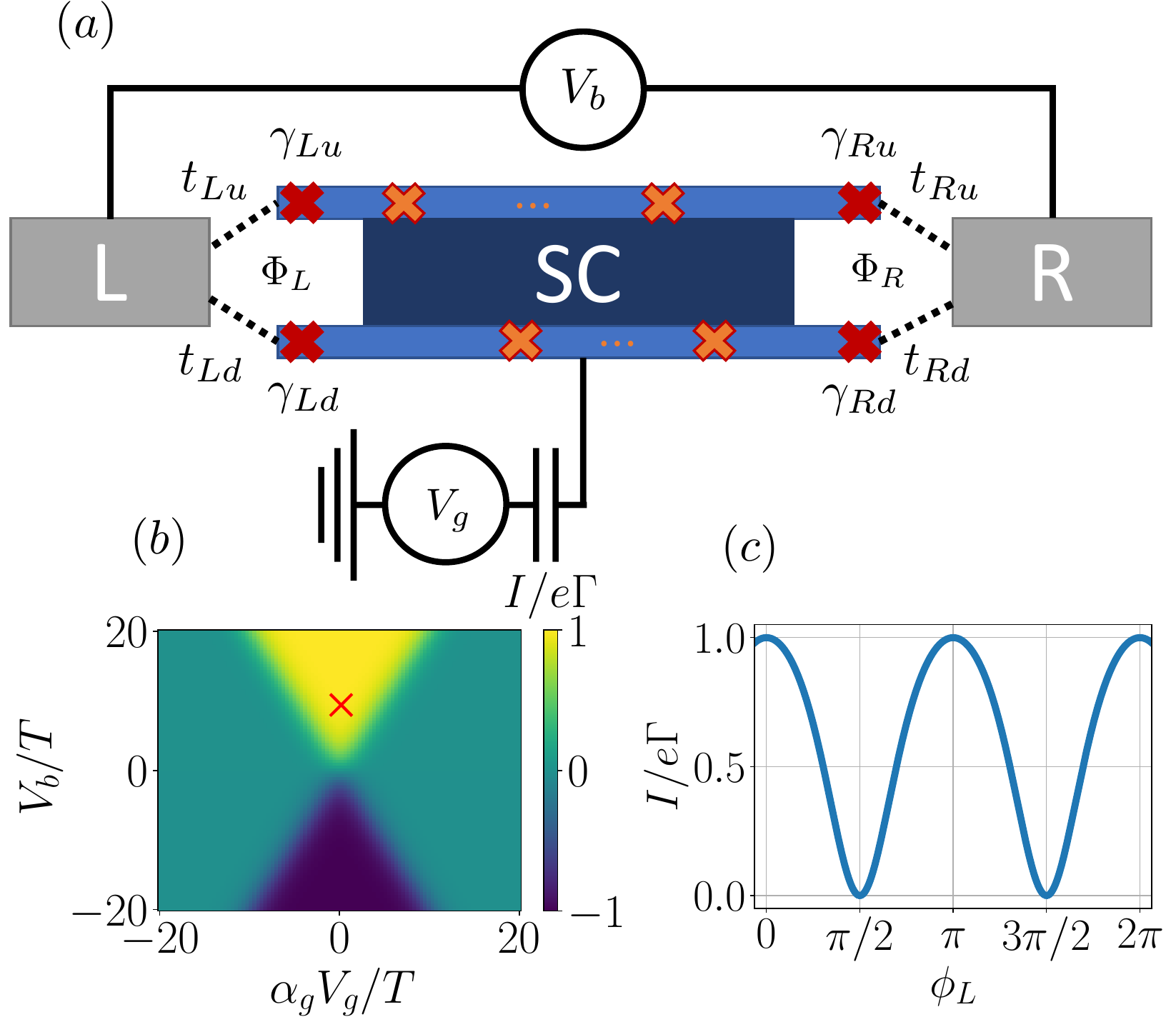}\\
		\caption{(a) Sketch of Majorana box, where two topological superconducting wires (blue), connected by a nontopological superconductor (dark blue), host four end MBSs (red crosses, operators $\gamma_{rm}$). Disorder might lead to additional unwanted MBSs (orange crosses). The box is tunnel coupled (amplitudes $t_{rm}$) to two normal leads ($L$ and $R$), subject to a voltage bias $V_b$. Magnetic fluxes $\Phi_L, \, \Phi_R$ are threaded through the loops associated with leads $L, \, R$ and cause relative phase differences $\phi_L, \, \phi_R$ between $t_{Lu}$ and $t_{Ld}$, $t_{Ru}$ and $t_{Rd}$. A gate voltage $V_g$ controls the equilibrium number of electrons on the Majorana box. (b) Current $I$ through the Majorana box as a function of $V_b$ and $V_g$ with the remaining parameters specified in the text. (c) $I$ at $V_g=0$~V, $V_b = 10 \, T$ (red cross in (a)) as a function of $\phi_L$ at $\phi_R = 0$.}
		\label{fig:setup}
	\end{center}
\end{figure}

In this work, we develop and employ a quantum master equation approach to investigate charge transport through a Majorana box where the source and drain contacts couple to two MBSs each, see Fig.~\ref{fig:setup}(a). We show that the same mechanism that allows quantum-dot-based parity readout \cite{munk2020parity,steiner2020readout,Schulenborg_PRB2021} induces a parity blockade in our transport setup, where the current is quenched and the qubit is stuck in a well-defined state. This is in close analogy to the Pauli spin blockade in double quantum dot spin qubits \cite{ono2002current,Hanson2007spins}. Just like the Pauli spin blockade, parity blockade can simplify various important qubit experiments. Fast and high-fidelity qubit initialization can be achieved by driving a current through the Majorana box which quickly gets stuck in the blocked state. The same principle can be used for readout, by applying a bias voltage such that an electron tunnels if the system is not in the blocking state. Single-shot readout can then be accomplished by charge detection on the box. Alternatively, measuring the current resulting from repeated operations provides an averaged readout. We solve the quantum master equation analytically for the clean box with four end MBSs and numerically for a disordered Majorana box with additional unwanted MBSs or topologically trivial Andreev bound states (ABSs). We show that the qubit coherence time can be read off from the remnant steady-state current in the blocking regime. This measurement requires neither fast manipulation, nor fast readout, only a DC transport measurement. Finally, we explain how to distinguish the clean Majorana box from the disordered system with additional MBSs or ABSs inside the box.
%We believe that our work can, in particular, facilitate first-generation experiments on imperfect topological qubits, where short decoherence times or other experimental limitations complicate single-shot protocols.

\emph{Model and transport theory.}
We consider the Majorana box transport setup sketched in Fig.~\ref{fig:setup}(a). Two topological superconducting nanowires are connected by a conventional (nontopological) superconductor and are tunnel coupled to electrically biased normal source and drain contacts. The Hamiltonian is $H=H_{MB}+ H_{res} +H_T$. The Majorana box is described by ($e= \hbar = k_B = 1$)
\begin{align}\label{eq:HMB}
    H_{MB}=\sum_{m=u,d} \frac{i}{2} \varepsilon_m \gamma_{Lm}\gamma_{Rm} + E_C(N-n_g)^2+H_{MB}^\mathrm{dis},
\end{align}
where $\gamma_{rm}$ are MBS operators and $\varepsilon_m$ is the overlap between MBSs in the same wire (our results remain qualitatively the same in the presence of additional overlaps between MBSs in different wires), $E_C$ is the charging energy,  $N$ counts the number of electrons (including Cooper pairs) on the Majorana box, and $n_g$ is the background charge controlled by the gate voltage $V_g$, $n_g = \alpha_g V_g$ with gate lever arm $\alpha_g$. $H_{MB}^\mathrm{dis}$ describes a number of additional unwanted MBSs [orange crosses in Fig.~\ref{fig:setup}(a)] induced by disorder, which may overlap with each other and with the edge MBSs $\gamma_{rm}$, see specific examples below. In this low-energy Hamiltonian we neglect the quasiparticle states above the superconducting gap.

The lead Hamiltonian is $H_{res} =\sum_{r} H_r$, with $H_r = \sum_{k} \xi_{rk} c_{rk}^\dagger c_{rk}$, where the $c_{rk}^\dagger$ create spinless electrons in lead $r=L, \, R$ with energies $\xi_{rk}$. We assume the leads to remain in thermal equilibrium at temperature $T$ and chemical potential $\mu_{L,R} = \pm V_b/2$. The tunneling between leads and Majorana box is described by
\begin{align}\label{eq:HT}
    H_T = \sum_{rmk}\gamma_{rm}\left( t_{rm}c_{rk} - t_{rm}^* c_{rk}^\dagger \right) + H_{T}^\mathrm{dis},
\end{align}
with tunnel amplitudes $t_{rm}$ which we take to be energy independent (wideband limit). We include magnetic fluxes $\Phi_L, \, \Phi_R$ threaded through the loops formed by leads $L, \, R$ and the end MBSs [Fig.~\ref{fig:setup}(a)] by adding a phase $\phi_r=2 \pi \Phi_r/\Phi_0$ to the upper tunnel amplitude of the left and right leads, $t_{ru} = |t_{ru}| e^{i \phi_r}$, $t_{rd}=|t_{rd}|$, where $\Phi_0$ is the flux quantum. The amplitude for a tunneling-induced transition between two many-body eigenstates $a$ and $b$ of the Majorana box is related to the tunnel matrix element $T_r^{ab} = \sum_{m=u,d}\langle a | \gamma_{rm} |b\rangle$. The typical time-scale of electron tunneling is then given by the tunnel rates $\Gamma_r^{ab} = 2\pi \nu_r |T_r^{ab}|^2$, where we take the density of states $\nu_r$ to be energy-independent within the bandwidth chosen as $D=100 T$. Unless stated otherwise, we will throughout the paper consider all tunnel amplitudes and densities of states to be equal, $t_{rm}=t$ and $\nu_r = \nu$, and define $\Gamma=2\pi \nu |t|^2$. $H_{T}^\mathrm{dis}$ describes the tunnel coupling of the disorder-induced MBSs to the leads.

Let us comment on two model assumptions which will be important for the results and which place some constraints on an experimental realization. First, the way $\gamma_{ru}$ and $\gamma_{rd}$ couple to the same lead channel in Eq.~(\ref{eq:HT}) is only strictly correct for an effectively 1D lead, but is a good approximation whenever tunneling from $\gamma_{ru}$ and $\gamma_{rd}$ occur into points of the lead separated by less than the Fermi wavelength. Second, considering spinless lead electrons is valid either when the magnetic field needed to induce the topological superconducting phase has fully spin-polarized the lead electrons around the Fermi level, or when the spin directions associated with allowed tunneling into $\gamma_{ru}$ and $\gamma_{rd}$ are aligned \cite{kjaergaard2012majorana} (which is the case for two identical wires).  

\emph{Quantum master equations.}
We focus on the regime of weak tunneling, $\Gamma \ll T$, but  strong electron-electron interaction $E_C$. Then it is appropriate to use a quantum master equation for the reduced density matrix $\rho$ of the Majorana box:
\begin{equation}
    \partial_t \rho = -i [H_{MB}, \, \rho] + W \rho.
\end{equation}
The quantum master equation consists of a unitary time evolution determined by $H_{MB}$ and a dissipative part introduced by the attached leads. We diagonalize the Majorana box Hamiltonian in Eq.~(\ref{eq:HMB}) to obtain the many-body eigenstates $|a_i\rangle$ and solve the master equation for the stationary state reduced density matrix $\rho_{a_1 a_2}$ and current $I$, where tunneling is treated in leading-order perturbation theory. We emphasize that, because of the near-degenerate ground state, it is important to solve for the full nondiagonal density matrix. All results presented below are obtained within a 1st order von Neumann quantum master equation~\cite{Kirsanskas_CPC2017} (equivalent to real-time diagrammatics in 1st order~\cite{konig1997cotunneling, leijnse2008kinetic, schoeller2009perturbative}). We have cross-checked that other approximations~\cite{Nathan_PRB2020, Kirsanskas_PRB2018} provide similar results, details are given in the supplementary information (SI) \cite{SI}.

Because of the large $E_C$ we consider only two charge states, arbitrarily denoted by $N=0$ and $N=1$, corresponding to the total parity of the MBSs being even or odd. We tune $V_g$ such that the two parity sectors are almost degenerate. The density matrix is diagonal in total parity, and a term in the master equation describing an electron tunneling onto or out of the Majorana box connects the two parity sectors. 

\emph{Parity blockade.} 
We first consider the clean box with $H^\mathrm{dis}_{MB} = 0$ and $\varepsilon_{m} = 0$. The current $I$ as a function of $V_b$ and $V_g$ for $\phi_L = \phi_R =0$ [Fig.~\ref{fig:setup}(b)] shows the Coulomb blockade pattern characteristic of transport through quantum dots \cite{kouwenhoven2001few}. The current is finite for $V_b, V_g$ such that there are available electrons in one contact that can tunnel into the Majorana box ($N \rightarrow N+1$) and available empty states in the other contact that can accept electrons tunneling out of the box ($N+1 \rightarrow N$). Otherwise current is suppressed by charging effects (Coulomb blockade).

For the remainder of the paper, we fix the voltages within the conducting regime [at the point marked by the red cross in Fig.~\ref{fig:setup}(b). We now vary $\phi_L$, see Fig.~\ref{fig:setup}(c), and find $I(\phi_L) \propto \cos^2 \phi_L$, meaning that the current is blocked for $\phi_L = (2n+1)\frac{\pi}{2}, \, n \in \mathrm{Z}$. To understand the blockade, we construct fermion operators using the two left and the two right MBSs: $f_L=(\gamma_{Lu}+i\gamma_{Ld})/2, \, f_R = ( \gamma_{Rd} + i \, \gamma_{Ru} )/2$. The eigenstates of the number operators $\hat{n}_r = f_r^\dagger f_r$, $r=L,R$, $|n_L n_R\rangle$, are also eigenstates of $H_{MB}$ when $\varepsilon_m = 0$ and $H_{MB}^\mathrm{dis} = 0$. Note that the even eigenstates, $|0_L 0_R\rangle$ and $|1_L 1_R\rangle$, are degenerate, and so are the odd eigenstates, $|0_L 1_R\rangle$ and $|1_L 0_R\rangle$. With this choice of basis and in this simple limit, the density matrix is diagonal. At the chosen voltages, electrons tunnel into the Majorana box from lead $L$ and out to lead $R$. For a current to flow, the state of the Majorana box must change according to $|0_L 0_R\rangle \rightarrow |1_L 0_R\rangle \rightarrow |1_L 1_R\rangle \rightarrow |0_L 1_R\rangle \rightarrow |0_L 0_R\rangle \rightarrow \hdots$ (electron tunnels in from the left, out to the right, in from the left, out to the right, $\hdots$). Note that because the number states $n_L,R$ are not charge eigenstates, it is possible to, for example, switch from $n_L=1$ to $n_L=0$ by an electron \emph{entering} the box from contact $L$.
Taking the tunneling term that adds an electron from the left lead in Eq.~(\ref{eq:HT}), and writing it in terms of the left/right fermion operator, we obtain: 
\begin{align}\label{eq:blockade}
H_{T,L} \rightarrow t \sum_k [c_k (e^{i \phi_L}+i)f_L^\dagger + c_k (e^{i \phi_L}-i)f_L]. 
\end{align}
For $\phi_L=\pi/2$ the second term in Eq.~(\ref{eq:blockade}) vanishes, which results in the transition $|1_L 1_R\rangle \rightarrow |0_L 1_R\rangle$ being suppressed. Therefore, the system becomes trapped in the blocking state $|1_L 1_R\rangle$ and no current can flow. For $\phi_L=3\pi/2$ the blocking state is instead $|0_L 0_R\rangle$. Reversing $V_b$ or changing $\phi_R$ at the right lead causes blocking instead in an odd state ($|0_L 1_R\rangle$ or $|1_L 0_R\rangle$). We note that, in direct analogy with the Pauli spin blockade \cite{ono2002current, danon2009pauli}, this parity blockade can be used for fast and high-fidelity initialization of a Majorana box qubit in any of the blocking states, as well as for readout in the corresponding basis. 
%The analogy with spin blockade becomes even clearer if we define a pseudospin within the two-level of each parity sector, such that the blocking states $|1_L 1_R\rangle$ ($|0_L 1_R\rangle$) corresponds to spin up ($s_z=1/2$) in the even (odd) sector; the blocking state then has a definite   

We now move on to investigate how the blockade is lifted and how to read off qubit lifetimes from the remnant current $I_{rem}=\min_{\phi_L}[I(\phi_L) ]$ in the blocked regime. First, we note that the blockade is lifted for asymmetric tunnel couplings to the upper/lower MBSs. We will quantify this more explicitly below and for now assume $t_{Lu}=t_{Ld}$. For now we keep the assumption $H_{MB}^\mathrm{dis} = 0$ but take $\varepsilon_m \neq 0$. Then the eigenstates are $|n_u n_d\rangle$ rather than $\ket{n_L n_R}$, associated with the up/down fermions with operators $f_m=(\gamma_{Lm}+i\gamma_{Rm})/2$ for $m=u,d$. The eigenenergies $E_{n_u n_d}$ within each parity sector are split by the MBS overlap, $2 \, \Delta_e = E_{11}-E_{00} = \varepsilon_u+\varepsilon_d$ and $2 \, \Delta_o = E_{01}-E_{10} = \varepsilon_u-\varepsilon_d$.  Moreover, the coupling to the leads introduces a Lamb-shift given by
\begin{equation}
\label{eq:lambshift}
H_{LS} = \Gamma I_P
    \begin{pmatrix}
    \sigma_x ( \sin \phi_L + \sin \phi_R ) & 0_2 \\
    0_2 & -\sigma_x ( \sin \phi_L - \sin \phi_R )
    \end{pmatrix}
\end{equation}
%which takes the form of additional overlap between the left and right Majorana pairs respectively
proportional to the principle value integrals $I_P$, see \cite{breuer2002theory,ptaszynski2019thermodynamics,SI}.

We can write the master equation in terms of the probability $p_{e/o}$ to be in the even/odd sector, and a pseudospin $\vec{s}_{e/o}$ that describes the density matrix within each sector, where we choose the $z$-axis to be along $|n_L n_R\rangle$. In the SI \cite{SI}, we derive Bloch-like equations for the pseudospin and show that the current is given by
\begin{equation}
I = 2e \, \Gamma (p_e + \sin \phi_L \, s_e^z ).
\end{equation}
Without MBS overlaps, $s_{e,o}^z$ are decoupled from $s_{e,o}^{x,y}$. Finite overlaps correspond to a magnetic field of strength $\Delta_{e,o}$ along the $x$-direction. In leading order perturbation theory, $\Delta_{e,o}$ induces an additional loss term of magnitude $\Delta^2/\Gamma^2 (1+I_P^2)$ in the master equation for $\partial_t s_e^z$ at $\phi_L = \pi/2$.
The blocking state corresponds to $p_e = 1- \frac{1}{2} \Delta^2/\Gamma^2 (1+I_P^2)$, $s_e^z = -1 + \Delta^2/\Gamma^2 (1+I_P^2)$, resulting in a current $I_\mathrm{rem}=e \Delta_e^2/\Gamma (1+I_P^2)$. This result can be generalized to any mechanism that allows parity to escape from the left Majorana pair ($n_L=0 \rightarrow n_L=1$) without changing the total charge on the Majorana box. If the parity escape rate is $\tilde{\Delta}/\hbar$, the resulting remnant current is $I_\mathrm{rem}=e \tilde{\Delta}^2/\Gamma (1+I_P^2)$ in the blocking regime. Interestingly, a larger tunnel coupling to the leads enhances the lifetime of the blocking state and suppresses current. Thus, even though a measurement of the remnant current directly gives the inverse lifetime of the blocking state $1/\tau = I_{rem}/e$, this is not the same as that for the isolated Majorana box. The Lamb-shift is not experimentally accessible and experiments can only extract the decay rate of the coupled Majorana box qubit $\tilde{\Delta}_\mathrm{coup} \equiv \tilde{\Delta}/(1+I_P^2)^{1/2}$ which is always smaller but of the same order of magnitude as the decay rate $\tilde{\Delta}$ of the isolated Majorana box. To measure $\tilde{\Delta}_\mathrm{coup}$ one should first extract $\Gamma$ from the current in the non-blocked regime, see Fig.~\ref{fig:setup}(c), and then measure $\tilde{\Delta}_\mathrm{coup}^2/\Gamma$ from the current in the blocked regime.
\begin{figure}[h]
	\begin{center}
		\includegraphics[angle=0,width=0.45\textwidth]{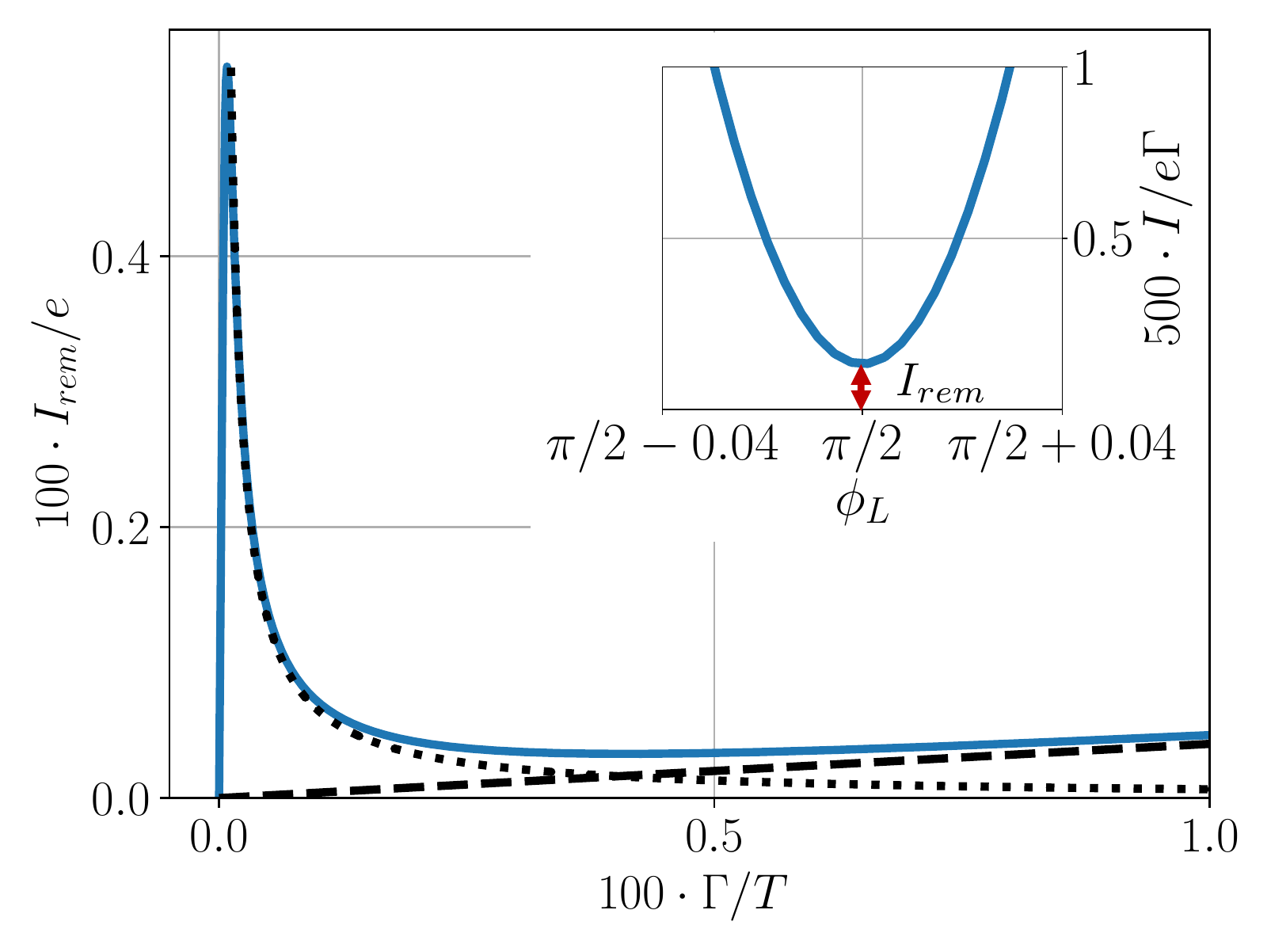}
		\caption{Inset: Lifting of the blockade at $\phi_L = \frac{\pi}{2}$ for $\varepsilon_{Lu} = 2 \cdot 10^{-4} \, T$ and $\delta_t = 10^{-2}$. The main plot shows $I_{rem}$ as a function of $\Gamma$ for $\delta_t = \delta_\phi = 10^{-2}$. After a sharp increase linear in $\Gamma$ until $\Gamma \approx \tilde{\Delta}_\mathrm{coup}$, $I_{rem}$ decreases as $e \, \tilde{\Delta}_\mathrm{coup}^2/\Gamma$ (black dotted line) before it increases again as $2e \, \delta^2 \Gamma$ (black dashed line). }
		\label{fig:remnantCurrent}
	\end{center}
\end{figure}

We illustrate this with a specific model for a disordered device containing unwanted MBSs. We assume that these MBSs are uncoupled to the leads, $H_{T}^\mathrm{dis}=0$, and
\begin{equation}
    H_{MB}^\mathrm{dis} = \frac{i}{2} \sum_{m=u,d}\sum_{r=L,R} \tilde{\varepsilon}_{rm} \gamma_{rm}\tilde{\gamma}_{rm}+\Omega_m \tilde{\gamma}_{Lm}\tilde{\gamma}_{Rm},
\end{equation}
where the $\tilde{\gamma}$s are four additional disorder-induced MBSs with couplings $\Omega_m$ between each other and couplings $\tilde{\varepsilon}_{rm}$ to the end MBSs. For $\tilde{\varepsilon}_{Lu} \gg \tilde{\varepsilon}_{Ld}, \, \Omega_m$, the relevant parity escape rate is $\tilde{\Delta} \approx (\tilde{\varepsilon}_{Lu} \pm \tilde{\varepsilon}_{Ld})/2 \approx \tilde{\varepsilon}_{Lu}/2$.
%But the escape rate becomes further suppressed for larger $\Omega_m$. Figure~\ref{fig:remnantCurrent}(b) shows the current in the blocking regime as a function of $\Gamma$.
Figure~\ref{fig:remnantCurrent} shows the current as a function of $\Gamma$. For $\Gamma \ll \tilde{\Delta}$ the current is proportional to $\Gamma$, it peaks at $\Gamma \approx \tilde{\Delta}$ and then decays with larger $\Gamma$ as $\tilde{\Delta}_\mathrm{coup}^2/\Gamma$ (dotted black line in Fig.~\ref{fig:remnantCurrent}). We also introduce deviations from the ideal blocked situation $\abs{t_{Ld}/t_{Lu} } = 1 - \delta_t, \, \phi_L = \frac{\pi}{2} + \delta_\phi, \, \delta = \sqrt{ \delta_t^2 + \delta_\phi^2 }$.
These deviations lead to a contribution to the current that is linear in $\Gamma$ (dashed black line) which dominates $I_{rem}$ for $\Gamma > \tilde{\Delta}_\mathrm{coup}/2 \delta$ \cite{SI}. In particular, the different scaling in $\Gamma$ makes it possible to distinguish experimentally between a remnant current caused by an escape rate ($I \propto 1/\Gamma$) compared to one due to finite $\delta$ ($I \propto \Gamma$).

%We now move on to a modified geometry in the form of a one-sided Majorana box, Fig.~\ref{fig:one-sided_box}(a), where three MBSs couple to the same lead and only a single MBS couples to the other lead. This has the advantage of alleviating the problem that the blockade is only complete for $t_{Lu} = t_{Ld}$. 

\emph{Distinguishing clean from disordered box.} 
Now we move on to showing that the phase dependence of the Lamb-shift offers the possibility to distinguish between the clean box with only four MBSs in total, and the disordered box with additional MBSs or ABSs, Fig.~\ref{fig:right-flux}(a). We model each ABS as two closely spaced MBSs which both couple to the leads by
\begin{equation}
    H_{T}^\mathrm{dis} = \sum_{rmk}\tilde{\gamma}_{rm}\left( t_{rm}c_{rk} - t_{rm}^* c_{rk}^\dagger \right),
\end{equation}
but with no overlaps with the MBSs on the other side of the box, $\Omega_m, \epsilon_U=0$. For each of these three cases, we block the current from the left lead with $\phi_L = \frac{\pi}{2}$ and investigate the dependence of $I_{rem}$ on $\phi_R$. Figure~\ref{fig:right-flux}(b) shows that the result is qualitatively different for the clean box (blue lines) compared with the disordered box (red lines) and ABS box (green lines), and this difference is robust to various parameter choices (different line styles).
\begin{figure}[h]
	\begin{center}
		\includegraphics[angle=0,width=0.45\textwidth]{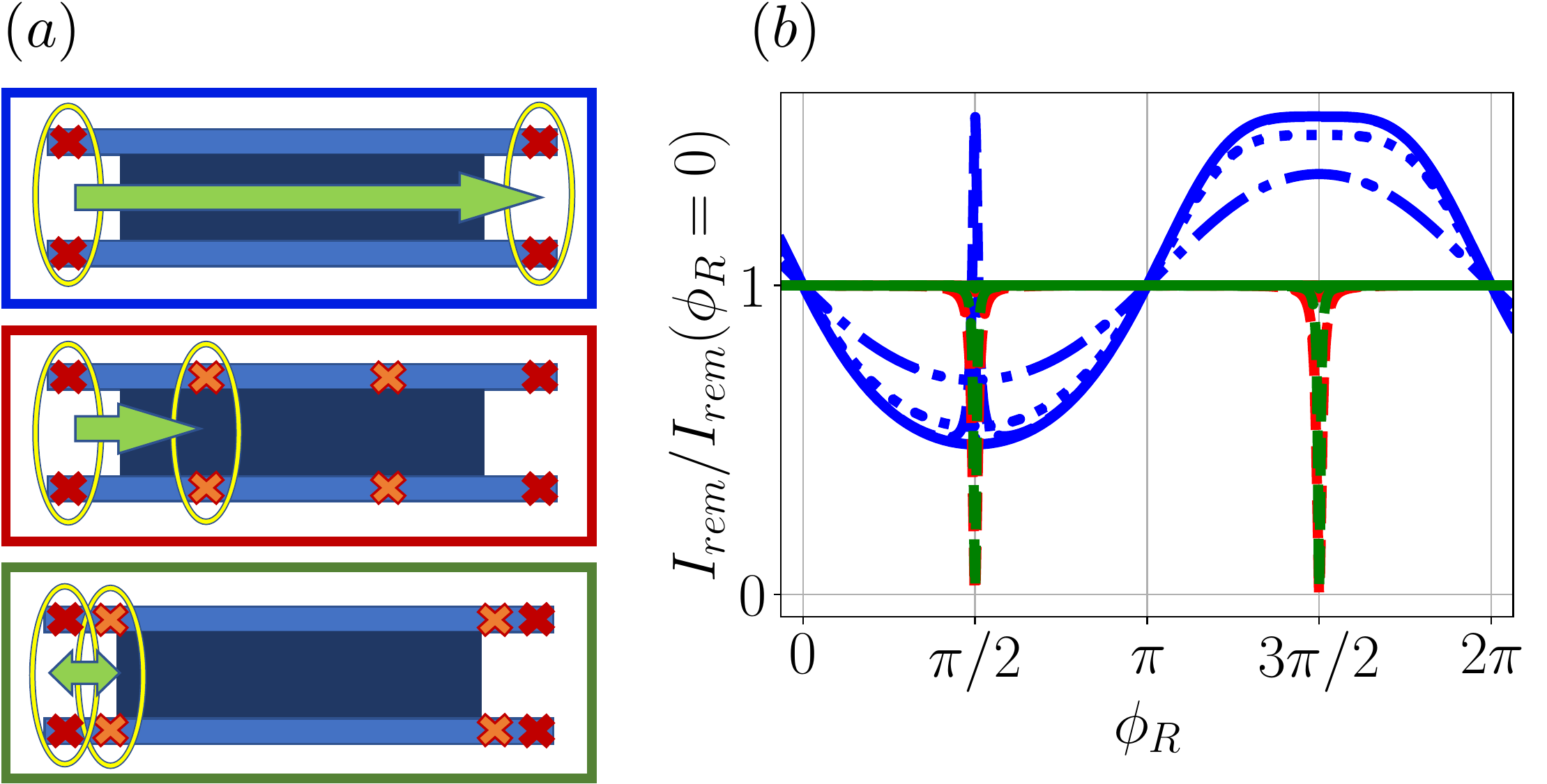}\\
		\caption{(a) Escape rate mechanisms leading to $I_{rem}$ for the clean box (blue), the disordered box (red) and the ABS box (green). (b) $I_{rem}$ as a function of $\phi_R$ at $\phi_L = \frac{\pi}{2}$ and $\Gamma = 10^{-2} \, T$ renormalized by $I_{rem}(\phi_R=0)$ for all three box models in line colours that match colours in (a). The overlaps are chosen as $\varepsilon_U, \varepsilon_{Lu} =10^{-5} \, T$ For each model we plot $I_{rem}$ for the perfectly fine-tuned setting (solid lines), deviation in the fine-tuned tunneling $\delta_t = 10^{-4}$ (dotted lines), mismatch in tunneling to left/right lead $\Gamma_R = \Gamma_L/2$ (dash-dotted lines) and a very large overlap $\varepsilon_U,\varepsilon_{Lu} =10^{-3} \, T$ (dashed lines). }
		\label{fig:right-flux}
	\end{center}
\end{figure}

In the case of the clean box the Lamb-shifts introduced by both leads either add up ($\phi_R=\frac{\pi}{2}$) or subtract ($\phi_R = \frac{3 \pi}{2}$), see Eq.~(\ref{eq:lambshift}). This leads to a decrease respectively increase in $I_{rem}$. This qualitative dependence is stable under all investigated parameter settings. For a large overlap, there appears a peak at $\phi_R = \frac{\pi}{2}$, which corresponds to additional blockade at the right lead interfering with the blockade at the left lead. %We find numerically that the width of this peak scales with $\tilde{\Delta}_\mathrm{coup}/\Gamma$.

For the disordered box with additional MBSs or ABSs, the line-shape is qualitatively different, with only very narrow dips (width $\propto \tilde{\Delta}_\mathrm{coup}/\Gamma$) around $\phi_R = \frac{\pi}{2}, \frac{3 \pi}{2}$. They mark a transition from a blockade at the left lead to a blockade at the right lead with smaller overlap/escape rate.

The qualitative difference of $I_{rem}(\phi_R)$ is due to the different escape mechanism from the parity blocked state to the next pair of MBSs, see Fig.~\ref{fig:right-flux}(a). In the clean box this nearest pair of MBSs is connected to the right lead. Accordingly, the adjustment of the Lamb-shift via $\phi_R$ affects the remnant current. But for the disordered and ABS boxes the nearest pair is located in the inner part/at the left end of the setup. There is no connection to the right lead and accordingly no dependence on the phase $\phi_R$. 
%One can say, that the lack of dependence results because the outer MBSs are shielded from on another by the inner MBSs.

\emph{Conclusions.}
We have used a quantum master equation approach to investigate transport through a Majorana box coupled to normal leads. There is a blocking regime, where the Majorana box becomes trapped in a well-defined state and the current is suppressed. In analogy with the Pauli spin blockade, this parity blockade can be used for qubit initialization and readout, as well as for measuring qubit coherence times from DC transport. We believe that this can become a key enabling technique for a first generation of MBS qubit experiments, where single shot readout might be challenging due to limited control or short qubit coherence times. Furthermore, the proposed setup makes it possible to experimentally distinguish between a clean Majorana box and a box with additional disorder-induced MBSs or ABSs. In our model, the parity lifetime is limited by MBS overlaps, but we expect that if quasiparticle poisoning is the dominant relaxation mechanism, the proposed measurement of the remnant current will instead reveal the poisoning time.
%We also envision future theoretical studies, for example investigating fusion-like protocols in the one-sided box, where cycles in the parameter space of the three couplings to the lower lead can be used for state preparation in one basis, followed by readout in another basis. 

\emph{Acknowledgments.}
%We can acknowledge people that have contributed in a positive way as well.
We acknowledge stimulating discussions with Karsten Flensberg, Michele Burello and Jens Schulenborg and funding from Nanolund, the Swedish Research Council (VR) and the European Research Council (ERC) under the European Union’s Horizon 2020 research and innovation programme under Grant Agreement No. 856526.

\bibliography{Interferometry}

\newpage\null\thispagestyle{empty}\newpage

\title{Supplementary information for interference and parity blockade in transport through a Majorana box}
\maketitle

In this Supplementary information (SI) we analytically derive the conditions for parity blockade for the clean box in a first order quantum master equation approach. We start by rewriting the quantum master equation into Lindblad form. Afterwards, we explain how the parity blockade is established without overlaps between the MBSs, $\varepsilon_u= \varepsilon_d=0,$ and with perfectly fine tuned tunnel couplings. Finally, we discuss how finite overlaps and deviation of the tunnel amplitudes lift this blockade.
\subsection{Transforming the first order von Neumann approach into Lindblad form}
The starting point of the derivation is the formulation of electron transport through a Majorana box in the 1st order von Neumann approach (1vN), see \cite{Kirsanskas_CPC2017}. It describes the time evolution of the quantum dot density matrix $\rho$
\begin{equation}
\begin{aligned}
\label{eq:1vNexact}
\partial_t \rho_{bb'} =& - i (E_b - E_{b'} ) \rho_{b b'}\\
&-i \sum_{b'', r} \rho_{b b''} \Big[ \sum_{a} \Gamma_{b''a, a b'}^{r} I_{ba}^{r-} - \sum_{c} \Gamma_{b''c, c b'}^{r} I_{cb}^{r+ *} \Big] \\
&-i \sum_{b'', r} \rho_{b'' b'} \Big[ \sum_{c} \Gamma_{bc, cb''}^{r} I_{cb'}^{r+} - \sum_{a} \Gamma_{ba, a b''}^{r} I_{b'a}^{r- *} \Big] \\
&-i \sum_{aa', r} \rho_{a a'} \Gamma_{ba, a'b'}^{r} \Big[ I_{b'a}^{r+*} - I_{ba'}^{r+} \Big] \\
&-i \sum_{cc', r} \rho_{cc'} \Gamma_{bc, c'b'}^{r} \Big[ I_{c'b}^{r-*} - I_{cb'}^{r-} \Big],
\end{aligned}
\end{equation}
where the first term describes the unitary time evolution of the quantum dot and the remaining ones describe the dissipative dynamics induced by the leads labeled by $r = L, R$. Depending on $N_b$ the indices $a, c$ sum over states with lower or higher total charge:
\begin{align}
N_a = N_b -1 && N_c = N_b + 1
\end{align}
The tunneling between leads and the Majorana box is described by the tunneling rate matrix $\Gamma$
\begin{equation}
\label{eq:gammaMatrix}
\Gamma_{ba, a'b'}^{r} = 2 \pi \nu_F \, T_{ba}^{r} \, T_{a'b'}^{r}.
\end{equation}
$I^{r\pm}$ contains the lead contribution 
\begin{equation}
I_{cb}^{r \pm} \equiv \frac{1}{2 \pi } \mathcal{P} \int_{-D_r}^{D_r} \frac{f \left( \pm \frac{E - \mu_r}{T_r} \right) }{E - E_{cb} } dE - \frac{i}{2} f( \pm x_{cb}^{r} ) \theta(D_r - \abs{E_{cb} } ),
\end{equation}
with the lead potential $\mu_r$, temperature $T_r$, bandwidth $D_r$, and the eigenenergies of the system $E_b$. We also define
\begin{align}
E_{cb} \equiv E_c - E_b, && x_{cb}^r \equiv \frac{E_{cb} - \mu_r}{T_r}.
\end{align}
Furthermore, we introduce the Fermi function $f(x)$ and the Heaviside step function $\theta(x)$
\begin{align}
    f(x) = [\exp(x) + 1]^{-1}, && \theta(x) = \begin{cases}
    0, \, x < 0 \\
    1, \, x \geq 0
    \end{cases},
\end{align}
as well as the principle value integral $\mathcal{P} \int_{-D_r}^{D_r} dE$. Throughout the SI, we focus on the red cross in Fig.~\ref{fig:setup}(b), were gate and bias voltage are chosen such that current can flu in principle. Furthermore, we assume the system and the leads to be in the limit
\begin{equation}
\label{eq:leadLim}
E_{cb} \ll T_r \ll \abs{\mu_r} \ll D_r,
\end{equation}
with the symmetries
\begin{align}
\label{eq:leadSym}
    T_L = T_R = T, && -\mu_R = \mu_L = \mu > 0, && D_L = D_R= D.
\end{align}
Due to the chosen limit and symmetries we neglect the eigenenergies compared to temperature, chemical potential and bandwidth by approximating
\begin{equation}
    \frac{+E_{cb} - \mu }{T} \approx - \frac{\mu }{T}
\end{equation}
within the principle value integrals, i.e. we drop the indices
\begin{equation}
\begin{aligned}
\label{eq:simplified}
I^{r \pm} \approx& \frac{1}{2 \pi } \mathcal{P} \int_{-D}^{D} \frac{f \left( \pm \frac{E - \mu_r}{T_r} \right) }{E} dE - \frac{i}{2} f( \pm x^{r} ),\\
 =& \frac{1}{2} \, I_P - \frac{i}{2} f( \pm x^{r} ), \\
    I_P \equiv& \frac{1}{\pi } \mathcal{P} \int_{-D}^{D} \frac{f \left( \frac{-E+\mu}{T} \right) }{E} dE, \hspace{1cm} x^r \equiv - \frac{\mu_r}{T_r},
\end{aligned}
\end{equation}
and split $I^{r \pm}$ into the principle value integral $I_P$ and the Fermi function. Note that the lead symmetries, Eq.~(\ref{eq:leadSym}), allow us to drop the index $r$ on the principle value integral. The approximation enables us to reformulate Eq.~(\ref{eq:1vNexact}) as
\begin{equation}
\label{eq:1vN}
\begin{aligned}
\partial_t \rho_{bb'} = &- i (E_b - E_{b'} ) \rho_{b b'}\\
&-i \sum_{b'', r=L, R} \rho_{b b''} \Big[ \sum_{a} \Gamma_{b''a, a b'}^{r} I^{r-} - \sum_{c} \Gamma_{b''c, c b'}^{r} I^{r+ *} \Big] \\
&-i \sum_{b'', r=L, R} \rho_{b'' b'} \Big[ \sum_{c} \Gamma_{bc, cb''}^{r} I^{r+} - \sum_{a} \Gamma_{ba, a b''}^{r} I^{r- *} \Big] \\
&-i \sum_{aa', r=L, R} \rho_{a a'} \Gamma_{ba, a'b'}^{r} \Big[ I^{r+*} - I^{r+} \Big] \\
&-i \sum_{cc', r=L, R} \rho_{cc'} \Gamma_{bc, c'b'}^{r} \Big[ I^{r-*} - I^{r-} \Big]
\end{aligned}
\end{equation}
These approximations allow us to reformulate the 1vN quantum master equation into a Lindblad form in the next section. We will start with defining matrices that describe the jump operators of the Lindblad master equation and the Lamb-shift.
\subsubsection{Definition of jump operators and Lamb-shift}
To obtain a Linblad form corresponding to Refs. \cite{Kirsanskas_CPC2017, Nathan_PRB2020}
\begin{equation}
    \partial_t \rho = -i [H, \rho] + L \rho L^\dagger - \frac{1}{2} \{ L^\dagger L, \rho \}
\end{equation}
we introduce the jump operators
\begin{equation}
\label{eq:jumpOp}
L_{kl}^r = T_{kl}^r \frac{1}{t_r}
\begin{cases}
\sqrt{f(-x^r)} \text{ if } N_k < N_l \\
\sqrt{f(+x^r)} \text{ if } N_k > N_l.
\end{cases}
\end{equation}
The operators absorb the Fermi function into the tunnel matrix. Furthermore, we use this definition to identify the tunneling amplitude $t_r$, defining the tunneling rate
\begin{equation}
\label{eq:gammaScalar}
    \Gamma_r \equiv 2 \pi \nu_F \, \abs{t_r}^2,
\end{equation}
which is assumed to be symmetric $\Gamma_R = \Gamma_L = \Gamma$ and smaller than the temperature but larger than the energies $E_{ab} \ll \Gamma_r \ll T$. We introduce the Lamb-shift as
\begin{align}
\label{eq:lambShiftDef}
H_{LS, kl} =& \pi \nu_F I_P \sum_{r, h} T_{k, h}^r T_{h, l}^r \cdot
\begin{cases}
+1 \text{ if } N_k < N_h \\
-1 \text{ if } N_k > N_h \\
\end{cases},
\end{align}
to absorb the unitary evolution introduced by the coupled leads. From the hermiticity of the tunnel matrices $T^r$, it is straightforward to show that $H$ is hermitian
\begin{equation}
    H_{LS, kl}^* = H_{LS, lk}.
\end{equation}
In the following section we use these definitions to rewrite the dissipation terms of Eq.~(\ref{eq:1vN}).

\subsubsection{Redefinition of the dissipation terms of the quantum master equation}
The neglected indices allow us to simplify
\begin{equation}
\label{eq:noPrinciple}
I^{r \pm *} - I^{r \pm} = -2 i \Im(I^{r \pm} ) = i f( \pm x^{r} )
\end{equation}
which lifts the principle value integrals within the last two summands in Eq.~(\ref{eq:1vN})
\begin{equation}
\label{eq:lastDiffSummand}
\begin{aligned}
-i& \sum_{aa', r} \rho_{a a'} \Gamma_{ba, a'b'}^{r} \Big[ I^{r+*} - I^{r+} \Big] +\\
-i& \sum_{cc', r} \rho_{cc'} \Gamma_{bc, c'b'}^{r} \Big[ I^{r-*} - I^{r-} \Big]\\
&\overset{\ref{eq:noPrinciple} }{=} \sum_{aa', r} \rho_{a a'} \Gamma_{ba, a'b'}^{r} f( +x^{r} ) + \sum_{cc', r} \rho_{cc'} \Gamma_{bc, c'b'}^{r} f( - x^{r} ) \\
&\overset{\ref{eq:gammaMatrix} }{=} 2 \pi \nu_F \bigg[ \sum_{aa', r} \rho_{a a'} T_{ba}^{r} \, T_{a'b'}^{r} f( +x^{r} ) + \\
  & \hspace{1.5cm}  + \sum_{cc', r} \rho_{cc'} T_{bc}^{r} \, T_{c'b'}^{r} f( -x^{r} ) \bigg] \\
&\overset{\ref{eq:jumpOp}}{=} 2 \pi \nu_F \bigg[ \sum_{aa', r} \abs{t_r}^2 L_{ba}^{r} \, \rho_{a a'} \, (L^{r\dagger})_{a'b'}  + \\
  & \hspace{1.5cm} + \sum_{cc', r} L_{bc}^{r} \, \rho_{cc'} \, (L^{r\dagger})_{c'b'} \bigg] \\
&= 2 \pi \nu_F \cdot \sum_{r} \abs{t_r}^2 (L^{r} \, \rho \, L^{r\dagger})_{bb'} \\
&\overset{\ref{eq:gammaScalar} }{=} \sum_{r} \Gamma_r \cdot (L^{r} \, \rho \, L^{r\dagger})_{bb'}.
\end{aligned}
\end{equation}
Next is the first dissipation term for which the Fermi function and the principle value integral in $I^{r \pm}$ both yield a non-zero contribution. We find
\onecolumngrid
\begin{equation}
\label{eq:firstDiffSummand}
\begin{aligned}
-i \sum_{b'', r} &\rho_{b b''} \Big[ \sum_{a} \Gamma_{b''a, a b'}^{r} I^{r-} - \sum_{c} \Gamma_{b''c, c b'}^{r} I^{r+ *} \Big] \\
\overset{\ref{eq:gammaMatrix}}{=} &-i \, 2 \pi \nu_F \, \sum_{b'', r} \rho_{b b''} \Big[ \sum_{a} T_{b''a}^{r} \, T_{ab'}^{r} I^{r-} - \sum_{c} T_{b''c}^{r} \, T_{cb'}^{r} I^{r+ *} \Big] \\
\overset{\ref{eq:simplified} }{=} &-i \, \pi \nu_F \, \sum_{b'', r} \rho_{b b''} \Big[ \sum_{a} T_{b''a}^{r} \, T_{ab'}^{r} I_P - \sum_{c} T_{b''c}^{r} \, T_{cb'}^{r} I_P \Big]
- \frac{1}{2}  \, 2 \pi \nu_F \, \sum_{b'', r} \rho_{b b''} \Big[ \sum_{a} T_{b''a}^{r} \, T_{ab'}^{r} f(-x^{r} ) + \sum_{c} T_{b''c}^{r} \, T_{cb'}^{r} f(x^{r} ) \Big] \\
\overset{\ref{eq:jumpOp} }{=} &-i \, \pi \nu_F \, \sum_{b'', r} \rho_{b b''} \Big[ \sum_{a} T_{b''a}^{r} \, T_{ab'}^{r} I_P - \sum_{c} T_{b''c}^{r} \, T_{cb'}^{r} I_P \Big]
- \frac{1}{2} \, 2 \pi \nu_F \, \sum_{b'', r} \abs{t_r}^2 \, \rho_{b b''} \Big[ \sum_{a} (L^{r\dagger})_{b''a} \, L_{ab'}^{r} + \sum_{c} (L^{r\dagger})_{b''c} \, L_{cb'}^{r} \Big] \\
\overset{\ref{eq:lambShiftDef} }{=} &+i \sum_{b''} \rho_{b b''} H_{LS, b'' b'}
- \frac{1}{2} \, 2 \pi \nu_F \, \sum_{b'', r} \abs{t_r}^2 \, \rho_{b b''} \Big[ \sum_{a} (L^{r\dagger})_{b''a} \, L_{ab'}^{r} + \sum_{c} (L^{r\dagger})_{b''c} \, L_{cb'}^{r} \Big] \\
\overset{\ref{eq:gammaScalar} }{=} &+i \, \sum_{b''} \rho_{b b''} H_{LS, b'' b'} - \frac{1}{2} \, \sum_{b'', r} \Gamma_{r} \, \rho_{b b''} \Big[ \sum_{a} (L^{r\dagger})_{b''a} \, L_{ab'}^{r} + \sum_{c} (L^{r\dagger})_{b''c} \, L_{cb'}^{r} \Big] \\
=&+i \, (\rho H)_{b, b'} - \frac{1}{2} \, \sum_{r} \Gamma_{r} \, (\rho \, L^{r\dagger} \, L^{r} )_{b, b'}
\end{aligned}
\end{equation}
\twocolumngrid
With the same steps as in Eq.~(\ref{eq:firstDiffSummand}), we find for the last remaining summand of Eq.~(\ref{eq:1vN})
\begin{equation}
\label{eq:secondDiffSummand}
\begin{aligned}
-i& \sum_{b'', r} \rho_{b'' b'} \Big[ \sum_{c} \Gamma_{bc, cb''}^{r} I^{r+} - \sum_{a} \Gamma_{ba, a b''}^{r} I^{r- *} \Big] \\
=&-i \, (H \, \rho )_{b, b'} - \frac{1}{2} \, \sum_{r} \Gamma_{r} \, (L^{r\dagger} \, L^{r} \, \rho )_{b, b'}
\end{aligned}
\end{equation}
Finally, we substitute Eqs. (\ref{eq:lastDiffSummand}), (\ref{eq:firstDiffSummand}) and (\ref{eq:secondDiffSummand}) into Eq. (\ref{eq:1vN}) and obtain
\begin{equation}
\label{eq:Lindblad}
\begin{aligned}
\partial_t \rho_{bb'} = &- i (E_b - E_{b'} ) \rho_{b b'} + i (H \, \rho)_{b b'} -i \, (\rho \, H)_{b'', b'} \\
&+ \sum_{r} \Gamma_r \cdot (L^{r} \, \rho \, L^{r\dagger})_{bb'} \\
&- \frac{1}{2} \Gamma_{r} \, (\rho \, L^{r\dagger} \, L^{r} )_{b, b'} - \frac{1}{2} \, \sum_{r} \Gamma_{r} \, (L^{r\dagger} \, L^{r} \, \rho )_{b, b'} \\
=& -i \left[H_{MB} + H_{LS}, \, \rho \right]_{bb'} \\
&+ \Gamma \sum_{r} \left( L^{r} \, \rho \, L^{r\dagger} - \frac{1}{2} \{ \rho, \, L^{r\dagger} \, L^{r} \} \right)_{bb'}.
\end{aligned}
\end{equation}
In the next section follows an explicit representation of the jump operators and the Lamb-shift for the clean Majorana box.

\subsection{Lindblad master equation of the clean box}
Without any overlaps, we are free to choose the basis in which we combine the MBSs without obtaining a non-diagonal Hamiltonian for the Majorana box. For simplicity, we combine the left and the right MBSs
\begin{align}
f_L = \frac{1}{2} ( \gamma_{Lu} + i \, \gamma_{Ld} ), && f_R = \frac{1}{2} ( \gamma_{Rd} + i \, \gamma_{Ru} ).
\end{align}
This defines the Fock states of the system as
\begin{align}
\label{eq:FS}
\ket{n_L, n_R}, && \mathcal{F} = \{ \ket{00}, \, \ket{11}, \, \ket{10}, \, \ket{01} \},
\end{align}
where the choice $\ket{11} = f_L^\dagger f_R^\dagger \ket{00}$ fixes the phase. Next, we use the fermionic operators $f_L, \, f_R$ to rewrite the tunnel Hamiltonian, Eq.~(\ref{eq:HT}), into
\begin{equation}
\begin{aligned}
H_{T} = t \sum_k \bigg ( &c_{L,k} \, [(e^{i \phi_L}+i)f_L^\dagger + (e^{i \phi_L}-i)f_L] \\
	& c_{R,k} \, [(e^{i \phi_R}+i)f_R^\dagger + (e^{i \phi_R}-i)f_R] \bigg )
	+ h.c.
\end{aligned}
\end{equation}
From this we obtain the tunnel matrices defined in Eq.~(\ref{eq:tunnelMatrices}).

\subsubsection{Evaluation of jump operators}
Using the tunnel matrices, we evaluate the jump operators and the Lamb-shift. Because of our assumption that the chemical potential is far bigger than the temperature of the leads, $\mu \gg k_B T$, we approximate the Fermi functions of Eq.~(\ref{eq:jumpOp}) to 0/1 depending on the sign of $x^r$. The physical interpretation for this is that the electrons are able to tunnel from the left lead into the box and from the box into the right lead but not the other way around. We obtain
\newcommand{\bigzero}{\mbox{\normalfont\Large 0}}
\newcommand{\rvline}{\hspace*{-\arraycolsep}\vline\hspace*{-\arraycolsep}}
\onecolumngrid
\begin{equation}
\label{eq:tunnelMatrices}
    \begin{aligned}
T_L = t \begin{pmatrix}
  \bigzero_2 & &
  \begin{matrix}
e^{-i \phi_L}-i & 0 \\
 0 & e^{-i \phi_L}+i 
  \end{matrix}\\
  \begin{matrix}
e^{i \phi_L}+i & 0 \\
 0 & e^{i \phi_L}-i 
  \end{matrix}
  & & \bigzero_2
\end{pmatrix}, &&
 T_R = t \begin{pmatrix}
  \bigzero_2 & &
  \begin{matrix}
0 & -i e^{-i \phi_R} + 1 \\
i e^{-i \phi_R} + 1 & 0
  \end{matrix}\\
  \begin{matrix}
0 & -i e^{i \phi_R} + 1 \\
i e^{i \phi_R} +1 & 0
  \end{matrix}
  & & \bigzero_2
 \end{pmatrix}.
\end{aligned}
\end{equation}
\twocolumngrid
\begin{equation}
\label{eq:jumpOpMat}
\begin{aligned}
L_L &= \begin{pmatrix}
 0 & 0 & 0 & 0 \\
 0 & 0 & 0 & 0 \\
e^{+i \phi_L}+i & 0 & 0 & 0\\
 0 & e^{+i \phi_L}-i & 0 & 0\\
\end{pmatrix}, \\
L_R &= \begin{pmatrix}
 0 & 0 & 0 & -ie^{-i \phi_R} + 1 \\
 0 & 0 & ie^{-i \phi_R} + 1 & 0 \\
 0 & 0 & 0 & 0 \\
 0 & 0 & 0 & 0 \\
\end{pmatrix}.
\end{aligned}
\end{equation}
\vspace{0.2cm}
Apart from the dynamics, we also need the jump operators to evaluate the current
\begin{equation}
\label{eq:lindbladCurrent}
I = \mathrm{Tr} \big( L_L \, \rho \, L_L^\dagger \big),
\end{equation}
see \cite{Kirsanskas_CPC2017}.
\subsubsection{Evaluation of the Lamb-shift}
Inserting Eq.~(\ref{eq:tunnelMatrices}) into Eq.~(\ref{eq:lambShiftDef}) yields
\begin{equation}
\label{eq:lambShiftMat}
\begin{aligned}
    H_{LS} = & \Gamma I_P \bigg[ 2
    \begin{pmatrix}
    -I_2 & 0_2 \\
    0_2 & I_2
    \end{pmatrix} \\
    &+ \begin{pmatrix}
    \sigma_z ( \sin \phi_L + \sin \phi_R ) & 0_2 \\
    0_2 & -\sigma_z ( \sin \phi_L - \sin \phi_R )
    \end{pmatrix}
    \bigg]
\end{aligned}
\end{equation}
The first term of Eq.~(\ref{eq:lambShiftMat}) only introduces a constant energy splitting between the even and odd parities. It is of the order $\Gamma \ll \mu$, so we can safely neglect its effect on the dynamics. The second term of the Lamb-shift, Eq.~(\ref{eq:lambShiftMat}), is more significant. It introduces a $\phi_L, \phi_R$ dependent rotation around the $z$-axis within each parity sector. Interestingly, this second term takes the form of additional overlaps between the left (right) pair of MBSs mediated via the leads. We can write it as 
\begin{equation}
    H_{LS} = \frac{i}{2} \varepsilon_L \, \gamma_{Lu} \gamma_{Ld} + \frac{i}{2} \varepsilon_R \, \gamma_{Ru} \gamma_{Rd}
\end{equation}
for the effective overlaps
\begin{align}
    \varepsilon_L \equiv -2\Gamma I_P \sin \phi_L, && \varepsilon_R \equiv 2\Gamma I_P \sin \phi_R.
\end{align}
In the next section, we insert these expressions into the Lindblad form and rewrite the dynamics into Bloch equations.

\subsection{Representation in Bloch equations}
Up to now, it was important to include both phases to understand how the contributions from the left and right leads to the Lamb-shift add up or subtract. In order to understand the blockade it is sufficient to consider one phase, so from now on we choose $\phi_R = 0$ such that $\varepsilon_R = 0$. We insert Eq.~(\ref{eq:jumpOpMat}) and Eq.~(\ref{eq:lambShiftMat}) into the Lindblad form in Eq. (\ref{eq:Lindblad}) and, after some matrix multiplication, we obtain a differential equation of the form
\begin{align}
\partial_t \rho = -i [H_{LS}, \rho] + 2 \, \Gamma ( - \rho + D ),
\end{align}
where the matrices $H_{LS}, D$ are defind in Eq.~(\ref{eq:MatD}).
Due to the total parity being a good quantum number the density matrix is block-diagonal
\begin{align}
\rho = \begin{pmatrix}
\rho_e & 0_{2}\\
0_{2} & \rho_o
\end{pmatrix}.
\end{align}
We separate it into sectors of even ($\rho_e$) and odd ($\rho_o$) total parity.
The last step of the derivation is to define the Bloch vectors for the even/odd parity sector $s_e/s_o$. They are chosen such that they fulfill
\begin{align}
\rho_e=\frac{p_e \, I_2 + \vec{s}_e \, \vec{\sigma} }{2}, && \rho_o=\frac{p_o \, I_2 + \vec{s}_o \, \vec{\sigma} }{2}.
\end{align}
In this notation $I_2$ is the identity and $\vec{\sigma}$ a vector of Pauli matrices. $p_e/p_o$ are the probabilities to measure the system in even/odd parity. Probability normalization gives
\begin{equation}
1= \mathrm{Tr}(\rho) = p_e+p_o.
\end{equation}
In total we have seven independent variables. The differential equation for the even probability and the $z$ component of the Bloch vectors in both sectors read
\begin{equation}
\label{eq:perfectBlockade}
\begin{aligned}
\partial_t p_e = 2 \, \Gamma (-p_e + p_o - \sin( \phi_L ) \, s_e^z ), \\
\partial_t s_e^z = 2 \, \Gamma (-s_e^z -s_o^z - \sin( \phi_L ) \, p_e ), \\
\partial_t s_o^z = 2 \, \Gamma (-s_o^z + s_e^z + \sin( \phi_L ) \, p_e ).\\
\end{aligned}
\vspace{0.2cm}
\end{equation}
\onecolumngrid
\begin{align}
\label{eq:MatD}
H_{LS} = \frac{1}{2} \begin{pmatrix}
 -\varepsilon_L & 0 & 0 & 0 \\
 0 & \varepsilon_L & 0 & 0 \\
 0 & 0 & \varepsilon_L & 0 \\
 0 & 0 & 0 & -\varepsilon_L
\end{pmatrix}, &&
D = \begin{pmatrix}
\rho_{33} - \sin(\phi_L) \rho_{00} & - i \rho_{32} & 0 & 0\\
i \rho_{23} & \rho_{22} + \sin(\phi_L) \rho_{11} & 0 & 0\\
0 & 0 & (1+\sin(\phi_L) ) \rho_{00} & i \cos(\phi_L) \rho_{01} \\
0 & 0 & -i \cos(\phi_L) \rho_{10} & (1-\sin(\phi_L) ) \rho_{11}\\
\end{pmatrix}
\end{align} 
\twocolumngrid
The equations are decoupled from the ones for the $x$ and $y$ components. Note that the $\sigma_z$ rotation of the Lamb-shift describes a unitary evolution within the $x-y$-plane. Accordingly, there is no contribution from it to Eq.~(\ref{eq:perfectBlockade}). 
Before we go on to the solution, we also need to express the current, specified in Eq.~(\ref{eq:lindbladCurrent}), in the Bloch representation as 
\begin{equation}
\label{eq:blochCurrent}
I  = 2 \, \Gamma (p_e + \sin \phi_L \, s_e^z ),
\end{equation}
which is also the version used in the main paper. Motivated by Evans theorem, which guarantees us always at least one zero eigenvalue for a Lindblad equation \cite{Evans1977irreducible, Evans1979generators, Manzano2020short}, we will now determine the stationary solution ($\partial_t \rho = 0$) for large times.

Let's first consider the case $\phi_L = 0$. It is easy to show that the only possibility for a stationary solution is given if both $z$ components are zero and the probability to find the system with even or odd parity are equally high
\begin{align}
\partial_t \rho = 0 && \Leftrightarrow && s_e^z = s_o^z = 0, \, p_e=p_o=\frac{1}{2},
\end{align}
leading to a current
\begin{equation}
I  = \Gamma.
\end{equation}
This is the unblocked case of electrons tunneling with rate $\Gamma$ through the device. We now turn to the more interesting blocking case and choose $\phi_L = \frac{\pi}{2}$ for which we find that
\begin{align}
\label{eq:blockedState}
\partial_t \rho = 0 && \Leftrightarrow && p_e = 1, \, s_e^z = -1, \, p_o = s_o^z = 0.
\end{align}
In this setting the only possibility for a stationary solution is that the even parity Bloch vector is aligned anti-parallel to the $z$-axis and the system is projected on the even parity sector. This results in $I  = 0$.
Indeed, a blockade is established. In the next section we will solve the system for finite overlaps between the MBSs and quantify the lifting of the blockade.

\subsection{Lifting of the parity blockade due to finite overlap}
We introduce small but finite overlaps between the MBSs
\begin{align}
0 < \varepsilon_u, \, \varepsilon_d \ll \Gamma.
\end{align}
From this we obtain the Hamiltonian of the system
\begin{align}
H_{MB} = \begin{pmatrix}
 -\Delta_e \, \sigma_x & 0_{2}\\
 0_{2} & \Delta_o \, \sigma_x
\end{pmatrix},
\end{align}
it introduces a rotation around the $x$-axis with the frequency
\begin{align}
    \Delta_e \equiv \frac{\varepsilon_u + \varepsilon_d}{2}, && \Delta_o \equiv \frac{\varepsilon_u - \varepsilon_d}{2},
\end{align}
for the even and odd parity sector respectively. These rotations couple the $z$-axis to the (previously uncoupled) $x-y$-plane. The corresponding unitary evolution reads
\begin{equation}
\begin{aligned}
-i [H_{MB}, \rho] =& -i \begin{pmatrix}
 -\Delta_e \, [\sigma_x, \rho_e] & 0_2\\
 0_2 & \Delta_o \, [\sigma_x, \rho_o]
\end{pmatrix} \\
 =& \begin{pmatrix}
 \Delta_e \, (s_e^z \sigma_y - s_e^y \sigma_z) & 0_{2}\\
 0_{2} & -\Delta_o \, (s_o^z \sigma_y - s_o^y \sigma_z)
\end{pmatrix}.
\end{aligned}
\end{equation}
Because of the overlaps we need to include all seven degrees of freedom to solve for the dynamics
\begin{equation}
\begin{aligned}
\partial_t p_e &= 2 \Gamma (1- 2 p_e - \sin \phi_L \, s_e^z ) \\
\partial_t s_e^z &= 2 \Gamma ( - s_e^z - s_o^z - \sin \phi_L  \, p_e - \xi \, s_e^y ) \\
\partial_t s_o^z &= 2 \Gamma ( - s_o^z + s_e^z + \sin \phi_L  \, p_e + \eta \, s_o^y ) \\
\partial_t s_e^x &= 2 \Gamma ( - s_e^x + s_o^y + \varepsilon_L/2\Gamma \, s_e^y) \\
\partial_t s_e^y &= 2 \Gamma ( - s_e^y + s_o^x - \varepsilon_L/2\Gamma \, s_e^x + \xi \, s_e^z ) \\
\partial_t s_o^x &= 2 \Gamma ( - s_o^x + \cos \phi_L \, s_e^y - \varepsilon_L/2\Gamma \, s_o^y) \\
\partial_t s_o^y &= 2 \Gamma ( - s_o^y - \cos \phi_L \, s_e^x + \varepsilon_L/2\Gamma \, s_o^x - \eta \,  s_o^z),
\end{aligned}
\end{equation}
with the small parameters
\begin{align}
\xi \equiv \frac{\Delta_e}{\Gamma}, && \eta \equiv \frac{\Delta_o}{\Gamma}.
\end{align}
As we are interested in the blockade lifting, a general solution to this equation is not needed. Instead, we solve this by expanding the solution for $\phi_L = \frac{\pi}{2}$ to leading order in $\eta, \, \xi$
\begin{align}
    s_o^y, s_o^x, s_e^y, s_e^x = 0 + O(\xi, \eta), && p_e = +1 + O(\xi, \eta), \\
    s_o^z = 0 + O(\xi, \eta), && s_e^z = -1 + O(\xi, \eta).
\end{align}
Setting $\phi_L = \frac{\pi}{2}$ also sets the Lamb-shift to $\varepsilon_L = -2\Gamma I_P$ and we obtain
\begin{equation}
\label{eq:finalEq}
\begin{aligned}
\partial_t p_e &= 2 \Gamma (1- 2 p_e - s_e^z ) \\
\partial_t s_e^z &= 2 \Gamma ( - s_e^z - s_o^z - p_e - \xi \, s_e^y ) \\
\partial_t s_o^z &= 2 \Gamma ( - s_o^z + s_e^z + p_e + \eta \, s_o^y ) \\
\partial_t s_e^x &= 2 \Gamma ( - s_e^x + s_o^y - I_P s_e^y ) \\
\partial_t s_e^y &= 2 \Gamma ( - s_e^y + s_o^x + I_P s_e^x + \xi \, s_e^z) \\
\partial_t s_o^x &= 2 \Gamma ( - s_o^x + I_P s_o^y) \\
\partial_t s_o^y &= 2 \Gamma ( - s_o^y - I_P s_o^x - \eta \,  s_o^z ).
\end{aligned}
\end{equation}
Inserting $s_o^z = 0 + O(\xi, \eta)$ into the last two lines of Eq.~(\ref{eq:finalEq}) we find that
\begin{align}
s_o^x, s_o^y = O(\eta^2, \xi \, \eta).
\end{align}
Therefore, we can neglect their contribution on the remaining set of equations. We find for the 5th and 6th line of Eq.~(\ref{eq:finalEq})
\begin{align}
\partial_t s_e^x &= 2 \Gamma ( - s_e^x - I_P s_e^y) \\
\partial_t s_e^y &= 2 \Gamma ( - s_e^y + I_P s_e^x + \xi \, s_e^z ),
\end{align}
which yields by inserting $s_e^z = -1 + O(\xi, \eta)$
\begin{align}
    s_e^y = \frac{-\xi}{1+I_P^2}, && s_e^x = \frac{-2 I_P \xi}{1+I_P^2}.
\end{align}
In turn we insert this into the 1st, 2nd and 3rd line of Eq.~(\ref{eq:finalEq}) and get
\begin{equation}
\begin{aligned}
\partial_t p_e &= 2 \Gamma (1- 2 p_e - s_e^z ), \\
\partial_t s_e^z &= 2 \Gamma ( - s_e^z - s_o^z - p_e + \frac{\xi^2}{1+I_P^2} ), \\
\partial_t s_o^z &= 2 \Gamma ( - s_o^z + s_e^z + p_e ). \\ \vspace{0.1cm}
\end{aligned}
\end{equation}
These equations are the same ones written in Eq.~(\ref{eq:perfectBlockade}) for the fully blocked system, with the effect of the overlaps included as an additional loss term $\frac{\xi^2}{1+I_P^2}$. We finally obtain the stationary state solution
\begin{equation}
\begin{aligned}
p_e = 1 - \frac{1}{2} \frac{\xi^2}{1+I_P^2}, && s_e^z = -1 + \frac{\xi^2}{1+I_P^2}, && s_o^z = \frac{1}{2} \frac{\xi^2}{1+I_P^2},
\end{aligned}
\end{equation}
which resembles the previous solution with a small misalignment $\propto \xi^2$ from the south pole of the Bloch sphere. Finally, we insert this into Eq.~(\ref{eq:blochCurrent}) and find for the remnant current
\begin{align}
I_{rem} = \Gamma \, \frac{\xi^2}{1+I_P^2} \overset{\xi = \Delta_e/\Gamma}{=} \frac{\Delta_e^2}{1+I_P^2} \, \frac{1}{\Gamma},
\end{align}
which is counter-intuitively proportional to $1/\Gamma$.
\subsection{Lifting of the parity blockade due to deviations in the tunnel coupling}
In this last section, we provide a short argument about the effect of a small deviation from the perfect blockade on the stationary current. For this we introduce the small deviation parameters $\delta_t, \delta_\phi \ll 1$ as
\begin{align}
t_{Ld} = (1 - \delta_t) \, t, && \phi_L = \frac{\pi}{2} + \delta_\phi, && \delta = \sqrt{\delta_t^2 + \delta_\phi^2}.
\end{align}
In a fine-tuned setting the blockade occurs because one of the tunneling matrix elements evaluates to zero
\begin{align}
    t_{Lu} + t_{Ld} = t \, (e^{i \phi_L} + i ) \overset{\delta_\phi = \delta_t = 0}{=} 0.
\end{align}
If we include finite deviations we obtain
\begin{equation}
\begin{aligned}
    t_{Lu} + t_{Ld} &= t \, e^{i \frac{\pi}{2} + \delta_\phi} + i t \, (1- \delta_t) \\
    &\approx t ( -\delta_\phi + i \delta_t),
\end{aligned}
\end{equation}
in leading order. We view this finite matrix element as a new tunnel coupling, which is added to the blocked dynamics. It defines the deviation tunneling rate as
\begin{align}
    \Gamma_{dev} = \abs{t ( -\delta_\phi + i \delta_t)}^2 = t^2 \delta^2.
\end{align}
As we are only interested in the leading order, we can neglect the effect of this additional tunnel coupling on the stationary state. We insert $\Gamma_{dev}$ into Eq.~(\ref{eq:blochCurrent}) for the stationary state from the perfect blockade Eq.~(\ref{eq:blockedState}) to obtain
\begin{align}
    I_{dev} = 2 \, \Gamma_{dev}.
\end{align}
This leading order contribution also holds if we include finite overlaps between the MBSs.

\end{document}